# Uncertainty-Aware Self-supervised Neural Network for Liver $T_{1\rho}$ Mapping with Relaxation Constraint

Chaoxing Huang,[1,5] Yurui Qian,[1] Simon Chun Ho Yu,[1,5] Jian Hou,[1] Baiyan Jiang,[1,4] Queenie Chan,[3] Vincent Wai-Sun Wong,[2] Winnie Chiu-Wing Chu,[1,5] Weitian Chen[1,5]

1. Department of Imaging and Interventional Radiology, The Chinese University of Hong Kong

2. Department of Medicine & Therapeutics, The Chinese University of Hong Kong

3. Philip Healthcare, Hong Kong SAR, China

4. Illuminatio Medical Technology Limited, Hong Kong SAR, China

5. CUHK Lab of AI in Radiology(CLAIR), Hong Kong SAR, China

Address correspondence to:

Weitian Chen

Room 15, Sir Yue Kong Pao Centre for Cancer

Prince of Wales Hospital Shatin, NT

Hong Kong SAR

(852)-3505-1036

Email: wtchen@cuhk.edu.hk

**Abstract**

*Objective*: $T_{1\rho}$ mapping is a promising quantitative MRI technique for the non-invasive assessment of tissue properties. Learning-based approaches can map $T_{1\rho}$ from a reduced number of $T_{1\rho}$ weighted images but requires significant amounts of high-quality training data. Moreover, existing methods do not provide the confidence level of the $T_{1\rho}$ estimation. We aim to develop a learning-based liver $T_{1\rho}$ approach that can map $T_{1\rho}$ with a reduced number of images and provide uncertainty estimation. *Approach:* We proposed a self-supervised learning neural network that learns a $T_{1\rho}$ mapping using the relaxation constraint in the learning process. Epistemic uncertainty and aleatoric uncertainty are modelled for the $T_{1\rho}$ quantification network to provide a Bayesian confidence estimation of the $T_{1\rho}$ mapping. The uncertainty estimation can also regularize the model to prevent it from learning imperfect data. *Main Results*: We conducted experiments on $T_{1\rho}$ data collected from 52 patients with non-alcoholic fatty liver disease. The results showed that when only collecting two $T_{1\rho}$-weighted images, our method outperformed the existing methods for $T_{1\rho}$ quantification of the liver. Our uncertainty estimation can further regularize the model to improve the performance of the model and it is consistent with the confidence level of liver $T_{1\rho}$ values. *Significance:* Our method demonstrates the potential for accelerating the $T_{1\rho}$ mapping of the liver by using a reduced number of images. It simultaneously provides uncertainty of $T_{1\rho}$ quantification which is desirable in clinical applications.

Keywords: quantitative MRI, $T_{1\rho}$ quantification, Self-supervised Learning, Uncertainty estimation,

## 1. INTRODUCTION

Quantitative MRI (qMRI) is a group of imaging techniques that measure tissue attributes from MR images based on known physical models (Keenan et al., 2019). It provides information that the conventional anatomical MRI cannot (Margaret Cheng et al., 2012). The quantification of spin-lattice relaxation time in rotating frame, known as $T_{1\rho}$, is a promising non-invasive technique for probing the biochemical properties of tissues. $T_{1\rho}$ has sensitivity to water content, metabolites, and macromolecules. It has been reported that $T_{1\rho}$ imaging is useful for assessment of chronic liver disease(Allkemper et al., 2014, Arihara et al., 2022, Chen et al., 2018, Takayama et al., 2015, Takayama et al., 2022, Wang et al., 2011, Xie et al., 2017) .

$T_{1\rho}$ is typically measured via spin-lock approaches. Traditionally, different durations of spin-lock pulse (or time of spin-lock, $TSL$) are chosen in the $T_{1\rho}$ preparation pulse sequence and multiple $T_{1\rho}$ weighted images (referred to acquired images or images in the following context) are acquired under different $TSL$s. These images comply an exponential decay model and are used to solve an inverse problem to fit the $T_{1\rho}$ map, as shown in **Fig. 1**. Note the scan time is proportional to the number of images acquired. Thus, it is desirable to quantify $T_{1\rho}$ values using a reduced number of acquired images. Deep convolutional neural networks have been used to reduce the scan time by k-space undersampling or reducing number of images (Feng et al., 2022). One limitation of deep learning is that it often requires a significant amount of high-quality training data, which is not always available in qMRI. Self-supervised learning was proposed to tackle this problem by using physics-based constraints to regularise the learning process and provide supervision signal from unlabelled data.(Grussu et al., 2021, Torop et al., 2020, Vasylechko et al., 2022, Huang et al., 2021). Many current deep leaning-based qMRI methods focus on the deterministic performance of the model, while ignoring the confidence (uncertainty) of the quantification results. The measurement of MRI parameters is inevitably uncertain because of the inherent randomness of the noise and imperfection of the data (Chen, 2015). For example, the presence of Rician noise can lead to overestimation of the relaxation values in a mono-exponential decay model (Raya et al., 2010) . The presence of $B_0$ and $B_1$ inhomogeneity can cause banding artifacts and quantification errors (Wáng et al., 2015, Chen et al., 2011, Chen, 2017) and thus the signal used for fitting can deviate from the ideal fitting model. Correction methods were developed to mitigate these effects but cannot completely remove the perturbations to the ideal exponential models. The uncertainty can also arise when the model is used to process samples that are distanced from the training data (like new lesion or abnormal anatomical structure), as the dataset is incomprehensive in covering all the real-world scenario. Therefore, it is crucial to provide uncertainty of quantification for adopting qMRI in clinical applications.

Uncertainty estimation has attracted significant interests in the community of learning-based qMRI recently. Glang et al.(Glang et al., 2020) calculated data uncertainty using a multi-layer perceptron to fit CEST z-spectra of multiple pools via a four-pool Lorentzian model. Similarly, Zhang et al. (Zhang et al., 2020) estimated the data uncertainty in susceptibility mapping. Qin et al.(Qin et al., 2021) proposed a Bayesian network for estimating a microstructure map using brain diffusion images. Shih et al. (Shih et al., 2021)evaluated the data uncertainty of fat quantification in the liver. All these methods use supervised learning. To the best of our knowledge, uncertainty in self-supervised qMRI has not yet been reported.

Previously, a self-supervised learning method for liver $T_{1\rho}$ mapping(Huang et al., 2021) was proposed by using a two-stream model. This approach, however, uses a relaxation model for self-supervised learning to fit the parametric map and the scaling coefficient map simultaneously, which is inherently ill-posed. In addition, directly learning from the imperfect data in liver imaging without proper regularisation may result in an erroneous gradient during learning. Furthermore, this method did not provide uncertainty estimation. This motivates us to develop a self-supervised model that learns the parametric map ($T_{1\rho}$ map) without explicitly learning the scaling coefficient map and simultaneously provides uncertainty estimation, which can in turn be used to regularise the model to improve its performance.

Given the above motivations, we explore a solution to better use the physics-based relaxation constraint for regularisation in the learning process, which may alleviate the need for a large amount of labelled training data. Our network is uncertainty-aware and can provide epistemic uncertainty and aleatoric uncertainty of $T_{1\rho}$ quantification. Epistemic uncertainty is related to the deficiency of data and knowledge and indicates the confidence of the predictions on new samples. Aleatoric uncertainty reflects the inherent data imperfection, and can regularise the model to prevent it from learning at areas where $T_{1\rho}$ data are imperfect. Our contributions are summarised as follows:

1. We propose a novel relaxation-constrained loss function for self-supervised learning in $T_{1\rho}$ mapping. We show that the proposed method outperformed the existing learning-based $T_{1\rho}$ quantification methods in liver $T_{1\rho}$ mapping on a non-alcoholic fatty liver disease dataset.

2. We model the uncertainty of the self-supervised learning network and propose a framework that can provide confidence on $T_{1\rho}$ quantification. We show that the uncertainty awareness further improves the performance of $T_{1\rho}$ quantification.

3. We analyse the resulting uncertainty maps and show that they reflect the reality in liver $T_{1\rho}$ imaging.

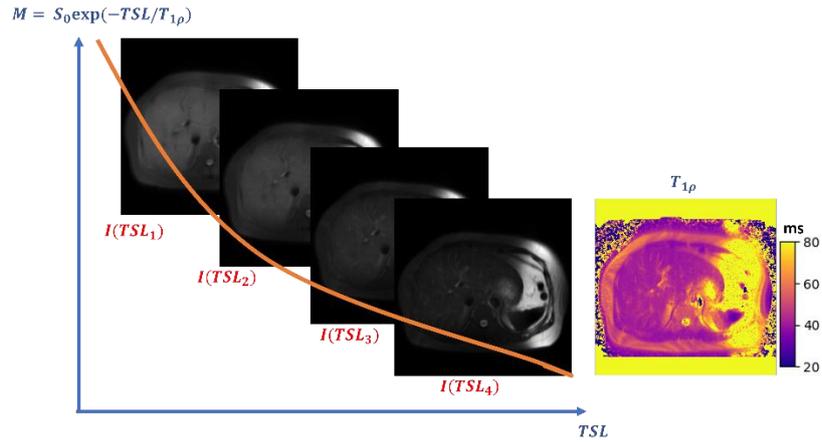

Figure 1. Fitting of $T1\rho$ quantitative map based on an exponential model. Multiple $T1\rho$ images need to be collected for a successful fitting. The setting of TSLs in our work is 0,10ms, 30ms, 50ms

## 2 Materials and Method

### 2.1 Dataset Acquisition and Dataset

The in vivo studies were conducted with the approval of the institutional review board. All MRI scans were conducted using a 3.0 T MRI scanner (Philips Achieva, Philips Healthcare, Best, Netherlands). A body coil was used as the RF transmitter, and with a 32-channel cardiac receiver coil (Invivo Corp, Gainesville, USA). We adopted a pulse sequence that can quantify $T_2$ and $T_{1\rho}$ simultaneously within a single breath-hold (Chen et al., 2017). Double inversion recovery combined with turbo spin echo acquisition was used for suppression of blood signal (Chen et al., 2016). The spectral attenuated inversion recovery (SPAIR) was applied for fat signal suppression. To reduce the magnetic field inhomogeneity, a pencil-beam volume shimming box was placed on the right lobe of the liver to reduce the $B_0$ field inhomogeneity (**Fig. S1** in the supplementary material ). The $B_1$ field inhomogeneity was reduced using dual transmit and vendor-provided RF shimming. The imaging parameters were as follows:

resolution = 1.5 × 1.5 $mm^2$, slice thickness = 7 mm, time of repetition = 2000 ms, and frequency of spin-lock = 400 Hz. The protocol acquired three slices of data from each subject. Four $T_{1\rho}$-weighted images were acquired from each slice within a single breath-hold with $TSL$ = 0, 10, 30, and 50 ms. The scan time to collect data for each slice was 10 s.

We retrospectively retrieved the data of 52 patients with non-alcoholic fatty liver disease. Our dataset has $52 \times 3 \times 4 = 624$ $T_{1\rho}$-weighted images and $52 \times 3 = 156$ $T_{1\rho}$ maps. Because each raw image, $I$, is in the DICOM format and contains the scaling information from the MR machine, and the scaling information is not relevant to the $T_{1\rho}$ property, images for the network input are normalised to the range from 0 to 1 and denoted as $\hat{I}$.

## 2.2 Inherent Data Imperfection in Liver $T_{1\rho}$ Imaging

Data imperfections are related to the acquisition protocol. Firstly, tissues, such as cerebrospinal fluid and residual blood which are not sufficiently suppressed during data acquisition, have a long $T_{1\rho}$. The maximum $TSL$ of 50ms used in the protocol, which is set based on the prior knowledge of our interested tissue (liver parenchyma), may not be able to capture sufficient relaxation signal of these tissues for their $T_{1\rho}$ quantification. Secondly, at regions where fat and blood signal are well suppressed, the images have low signal-to-noise ratio (SNR) which can lead to unreliable quantification. Thirdly, the spin-lock pulse cluster used for $T_{1\rho}$ imaging in this work can fail when $B_0$ and $B_1$ field inhomogeneities are significant. In the protocol for data acquisition, a shim box is placed on the right lobe of the liver (left part of the image) to ensure magnetic field homogeneity in the liver by sacrificing homogeneities in other areas such as the stomach. Thus, tissues other than liver parenchyma may have image artefacts because of inhomogeneous magnetic fields at those regions. Fourthly, the flow effect and motion may also contribute to the data imperfection.

To sum up, multiple factors including noise, $B_0$ and $B_1$ field inhomogeneities, and motion can contribute to the inherent data imperfectness in some parts of the image. Therefore, it is important to make the network uncertainty aware to improve the prediction performance.

**2.3 Algorithm design**

**2.3.1 Relaxation-constrained loss**

Given an on-resonance $T_{1\rho}$ relaxation model, the signal equation of an image can be written as:

$$I(x,y) = S_0(x,y)\exp\left(-\frac{TSL}{T_{1\rho}(x,y)}\right) \tag{1}$$

where $S_0$ is a scaling parameter which can be affected by hardware settings, acquisition parameters, and tissue properties. $(x,y)$ is the coordinate of a pixel. Conventionally, multiple images are acquired with different $TSL$, and they are used to fit $S_0$ and $T_{1\rho}$ at each location via least square fitting methods, as illustrated in **Fig. 1**. Theoretically, the $T_{1\rho}$ value can be computed by taking the logarithm of the quotient of two images with two different $TSL$s under the assumption that the images have sufficient SNR. In practice, directly applying this method is sensitive to noisy outlier and can produce erroneous values. On the other hand, this quotient relationship can be used as a constraint in self-supervised learning without explicitly learning the $S_0$. Let us denote a convolutional neural network that takes two normalised images under various $TSL$s as an input pair for mapping as $f(\cdot)$. The output of the network can be denoted as:

$$T_{1\rho}^{(i,j)} = f([\hat{I}(TSL_i), \hat{I}(TSL_j)]) \tag{2}$$

where $(i,j)$ is the index of the input pair. It should satisfy the constraint between an arbitrary pair of images under various $TSL$s from the same slice in the dataset as follows:

$$I(TSL_m) = I(TSL_n)\exp\left(\frac{TSL_n - TSL_m}{T_{1\rho}^{(i,j)}}\right) \qquad (3)$$

where $m, n$ are the indexes of the images in the pair for constraint. Since single pair of constraint (referred to constraint pair in the following context) is sensitive to noise, we apply this constraint to all possible constraint pairs in the same slice. The theoretical loss function can therefore be calculated as:

$$L = \frac{\sum_{s=1}^{S}\sum_{p_{\{m,n\}}=1}^{P2}\sum_{p_{\{i,j\}}=1}^{P1}\left|I(TSL_m) - I(TSL_n)\exp\left(\frac{TSL_n - TSL_m}{T_{1\rho}^{(i,j)}}\right)\right|}{P_1 * P_2 * S} \qquad (4)$$

where $P_1$ and $P_2$ are the number of input pairs and the number of constraint pairs in a slice, respectively. $p_{\{m,n\}}$ and $p_{\{i,j\}}$ stand for the index of one possible pair of constraint and the index of one possible pair of input, respectively. $S$ is the number of slices in the training set. In practice, we back-propagate the loss from every constraint pair separately by inputting one input pair into the network for multiple times for better convergence.

The above relaxation-constrained learning process eliminates the need for explicitly learning $S_0$. The model is also expected to learn a robust $T_{1\rho}$ map that satisfies the relaxation constraint between different pairs of images of different $TSL$s. The self-supervised learning pipeline is shown in **Fig. 2**.

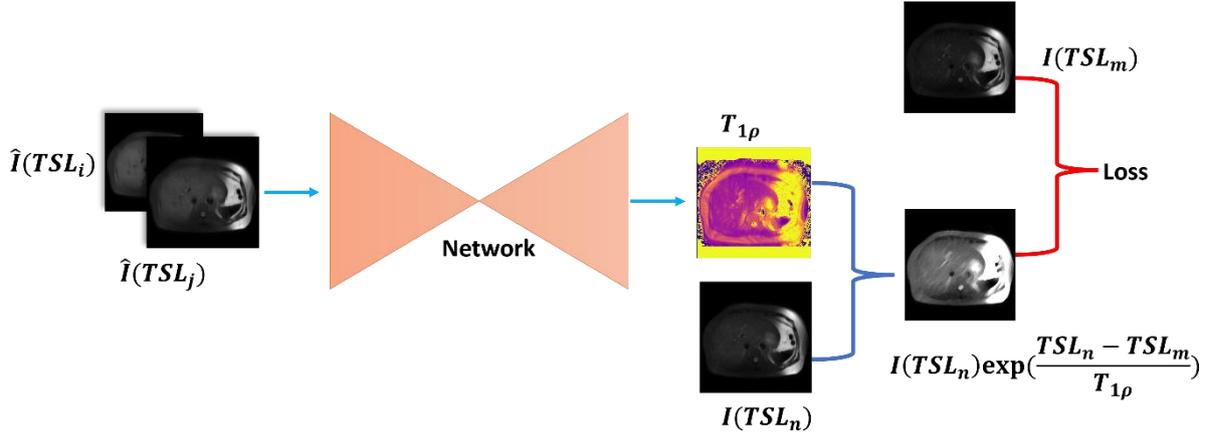

Figure 2: Self-supervised learning pipeline. By taking the input pair of a slice, the network produces a $T1\rho$ map that satisfies the relaxation constraint between two $T1\rho$-weighted images from the slice.

### 2.3.2 Modelling the uncertainty

We model the uncertainty following the Bayesian neural network in computer vision, with epistemic uncertainty and aleatoric uncertainty (Kendall and Gal, 2017, Hüllermeier and Waegeman, 2021).

**Epistemic Uncertainty:** The epistemic uncertainty is modelled by defining a distribution over the network's learnable parameters, and the performance variance between a set of sampled networks is measured. The epistemic uncertainty reflects how ignorant the model is given the training data. The epistemic uncertainty can be large if the training data do not represent a test sample well. We adopt Monte Carlo Dropout to approximate the epistemic uncertainty (Gal and Ghahramani, 2016). During testing, the dropout layers (Srivastava et al., 2014) in the network are enabled, and $K$ forward passes are performed to sample $K$ different models. The mean and the variance are computed as

$$\mu = \frac{1}{K}\sum_{k=1}^{K} T_{1\rho k} \qquad (5)$$

$$\Sigma = \frac{1}{K}\sum_{k=1}^{K}(T_{1\rho k} - \mu)^2 \tag{6}$$

Where $T_{1\rho k}$ is the predicted map of every sampled model.

**Aleatoric Uncertainty:** Aleatoric uncertainty reflects the inherent randomness of the data. If some of the patterns in a test sample are similar to the noisy patterns in the training set, they are likely to have high uncertainty. We model the output of our network as the parameters of a probability distribution over the $T_{1\rho}$ value consistent with Bayesian neural network predictive approaches. We denote the probability function of this distribution as $p(T_{1\rho}|f)$, where $f$ is the mapping network. We model the distribution as a Laplacian distribution parameterised by mean $\mu_L$ and standard deviation $\sigma_L$. By learning the model through log-likelihood maximisation, the loss function can be written as:

$$\frac{|\mu_L - \overline{T_{1\rho}}|}{\sigma_L} + \log(\sigma_L) \tag{7}$$

The ground-truth $T_{1\rho}$ value $\overline{T_{1\rho}}$ is absent, and we need to adapt the loss function in a self-supervised way. It should reflect the trend that large uncertainty value is associated with those patterns with noisy signal. Intuitively, we can assume that the pixels with large error in Eq [4] are related to pixels with inaccurate $T_{1\rho}$ predictions or noisy signals that violate the mono-exponential model. Therefore, the loss can be further modified as follows:

$$L = \frac{1}{P_1 P_2 S} \sum_{s=1}^{S}\sum_{p_{\{m,n\}}=1}^{P_2}\sum_{p_{\{i,j\}}=1}^{P_1}\left(\frac{\left|I(TSL_m) - I(TSL_n)\exp\left(\frac{TSL_n - TSL_m}{T_{1\rho}^{(i,j)}}\right)\right|}{\sigma_L} + \log(\sigma_L)\right) \tag{8}$$

The modelled aleatoric uncertainty map is learnt by the network. The intuition of minimising this loss function is straightforward. The level of uncertainty tends to increase if the residual is large, as a large residual indicates noisy or erroneous data that are difficult for the model to learn. A high level of uncertainty discourages the model from learning from patterns with a low confidence level. In practice, the network directly learns the logarithm of the uncertainty to avoid numerical instability.

**Bayesian Uncertainty**: After modelling the epistemic uncertainty and aleatoric uncertainty, we combine them together additively to form the Bayesian uncertainty:

$$Var = \frac{1}{K}\sum_{k=1}^{K}\left[\left(T_{1\rho k} - \mu\right)^2 + \sigma_{Lk}^2\right] \tag{9}$$

where $\sigma_{Lk}^2$ is the aleatoric uncertainty map obtained at the $k$th sampled network.

### 2.3.3 Network architecture

The network architecture is based on a UNet model (Ronneberger et al., 2015) used for brain abnormality segmentation in MRI (Buda et al., 2019). The $T_{1\rho}$ mapping value is closely associated with the tissue property. The semantic information between tissues can therefore benefit the feature extraction. The skip connection in the UNet may help extract semantic information of different tissues at different scales. We replace the original three-channel input layer of the UNet with a two-channel convolution layer and the original one-channel output layer with a two-channel convolutional layer. Dropout layers with a dropout rate of 0.25 are added in the decoder part. The architecture is shown in **Fig. 3**. The configuration of every convolutional block is shown in **Table 1**.

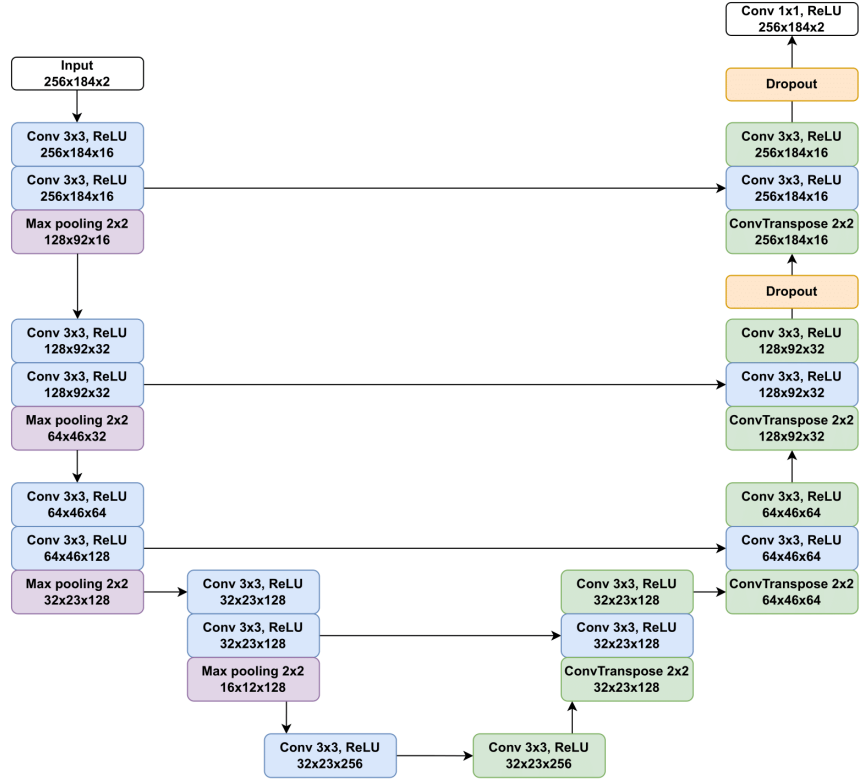

Figure 3: Network architecture

| Components | Order |
| --- | --- |
| Conv(kernel size =3,padding=1,stride=1) | 1 |
| Batch Normalisation | 2 |
| ReLU | 3 |

Table 1: Configuration of the convolutional block

### 2.3.4 Evaluation Metrics

**ROI Mean Absolute Error:** We followed the region of interest (ROI) based evaluation method in deep learning-based qMRI methods in liver and knee imaging(Shih et al., 2021, Sveinsson et al., 2021), by comparing the pixel-wise mean absolute error between the inference $T_{1\rho}$ map and the reference $T_{1\rho}$ map in the ROI for performance evaluation. The ROI Mean Absolute Error (**RMAE**) is expressed as:

$$RMAE = \frac{1}{N}\sum_{n=1}^{N}|T_{1\rho n} - \overline{T_{1\rho n}}|_{(T_{1\rho n}, \overline{T_{1\rho n}}) \in ROI} \qquad (10)$$

where $N$ is the total number of pixels within the ROI, $T_{1\rho n}$ and $\overline{T_{1\rho n}}$ are the predicted value and the reference value of a pixel respectively. The final result is reported by taking the mean of the RMAE of all the predicted maps in the test set. The ROI is manually drawn on the right-lobe of the liver to cover the parenchyma as much as possible while avoiding large vessels and bile-ducts. The drawing was performed on a $T_{1\rho}$ weighted image before $T_{1\rho}$ fitting to ensure the evaluation fairness. We used the $T_{1\rho}$ map fitted by four $T_{1\rho}$-weighted images using the non-linear least square fitting method as the reference map.

**ROI Mean Value Absolute Error:** It is common to calculate the mean $T_{1\rho}$ value of all the pixels in the ROI in pathological study to filter out the noisy values. This mean value is then used to represent the $T_{1\rho}$ value of the predicted map. Therefore, we first compute the mean value in the ROI of both the predicted map and the reference map, and then calculate the absolute error between these two. The ROI Mean Value Absolute Error (**RMVAE**) is expressed as:

$$RMVAE = |(\frac{1}{N}\sum_{n=1}^{N}T_{1\rho n}) - (\frac{1}{N}\sum_{n=1}^{N}\overline{T_{1\rho n}})|_{(T_{1\rho n}, \overline{T_{1\rho n}}) \in ROI} \qquad (11)$$

The final result is reported by taking the mean of the RMVAE of all the predicted maps in the test set. We also conduct the pair t-test($p<0.05$ as significant) to test the null hypothesis that there is no significant differences between the ROI mean value of the predicted maps from the proposed model and the reference maps.

**ROI Pixel Standard Deviation:** The predicted map is expected to reflect the uniform tissue property of the liver parenchyma. Thus, we calculate the standard deviation of the pixel-wise $T_{1\rho}$ value in the ROI of each predicted map. We refer it as ROI Pixel Standard Deviation

(**RPSD**). The final result is reported by taking the mean of the RPSD of all the predicted maps in the test set.

**ROI Structure Similarity (RSSIM):** We adopt the Structure Similarity(Wang et al., 2004) in the ROI as a metric to see if the predicted map can reveal the anatomical structure of the liver. The pixels in the ROI are included for SSIM computation. We refer it as ROI Structure Similarity (**RSSIM**). Note that the relaxation time information is more essential than the anatomical structure information in the quantitative map for biochemical study, and the SSIM serves only for auxiliary evaluation purpose.

**Sparsification Plot:** Sparsification plot is a common way of evaluating the uncertainty quantification(Ilg et al., 2018, Poggi et al., 2020). All the pixels in the ROI of the predicted $T_{1\rho}$ maps are ranked in descending order according to the uncertainty values. Iteratively, a subset of pixel (the top 5% in the ranked pixels were used in this study) was removed, and the RMAE value was computed for the remaining pixels. We can then get a curve of the RMAE according to the removed fraction. Two other reference curves were also computed. An ideal curve, termed the Oracle curve, was computed by directly ranking the pixels according to the RMAE value. The Oracle curve always has a descending trend. A random curve was computed by randomly omitting a subset of the pixels every time. This curve should have a flat trend since it does not tell any information. A reasonable uncertainty sparsification curve should have a descending trend to reflect the intuition that pixel predictions with a larger error have greater uncertainty.

## 3 Experiments and Results

### 3.1 Implementation Details

The data of the 52 patients were randomly divided into 4 groups, with 13 patients in each group. All the experiment results are based on the four-fold cross-validation scheme similar to those in (Nie et al., 2018) and (Bortsova et al., 2019), where each fold has its own test set. In each fold, one group of the 13 patients was used for testing and the rest of the 39 patients were used for training and validation. Among those 39 patients, we randomly chose 35 patients' data for training and 4 patients' data for validation. This procedure was repeated four times and the mean performance were reported as the result. Data augmentation was applied as follows: central rotation with degrees of -7.5, -5, -2.5,2.5,5,7.5; shift with pixel values of -10,-5,10 and 5 in four directions (vertical, horizontal and two diagonals). The input pairs are formed by an image with $TSL = 0$ and another image with a different $TSL$. The constraint pairs are all the possible pairs in the training set. All images are resized to $256 \times 256$. The learning rate was $5 \times 10^{-4}$ and the weight decay was 1e-4. Adam(Kingma and Ba, 2014) was used as the optimizer. The experiments were performed on Pytorch 1.9(Paszke et al., 2019), with a GTX 1080ti GPU. Each fold of the experiments cost around 70 epochs, and early stopping was applied. All the above settings remain unchanged unless specially indicated. The number of sampled models $K$ used to calculate uncertainty was set to 20, and every forward pass took approximately 0.016 seconds.

### 3.2 Comparison Experiments of Different Models

To examine the performance of our proposed model, we compared its performance with the following models:

**2-TSL**: The $T_{1\rho}$ map is computed by taking the logarithm of the quotient of the two input images. This model is simple but vulnerable to noise.

**Two-stream:** The implementation of this model is similar to that of the self-supervised learning model in previous study (Huang et al., 2021). This model requires learning $S_0$ and $T_{1\rho}$ map by using two branches of neural network, and the output is used to synthesis another image in the same slice. To ensure a fair comparison, we replaced the original input layers with a two-channel input convolutional layers to ensure that the input was the same as that in the proposed method. The input of the $S_0$ branch is the image with $TSL = 0$. Apart from these, the implementations and settings are the same as the original study.

**Supervised Learning**: This model uses the same network architecture in the proposed method and the network was trained in a fully supervised way. The only reference for supervision during training for every input pair is the $T_{1\rho}$ map fitted by four $T_{1\rho}$ images. The loss function is the L1 norm.

**Table. 2** shows the performance of different models. The proposed method outperforms the Two-steam method and the other methods regarding the general performance. While the proposed method has a marginally lower RSSIM than the Two-Stream baseline, the value with 0.9264 already demonstrates a decent anatomy similarity in the ROI.

**Fig. 4** shows the predicted $T_{1\rho}$ maps using different methods. Note the proposed method produces comparable results to the reference map. The results of 2-TSL and the supervised learning are inferior. The $T_{1\rho}$ map from the 2-TSL approach is noisy with an overall lower $T_{1\rho}$ value, and this is aligned with the high RPSD result in Table 2. The maps produced by the supervised learning are poor at revealing the anatomical structures with over-smoothing visual effect. Detailed discussion is provided in the discussion section.

| Models | RMAE(ms) | RPSD(ms) | RMVAE(ms) | RSSIM |
|---|---|---|---|---|
| 2-TSL | 7.55 | 39.15 | 4.95 | 0.7088 |
| Two-Stream | 4.28 | 8.00 | 3.61 | 0.9285 |
| Supervised | 5.82 | 7.37 | 5.13 | 0.8851 |
| Proposed | 3.60 | 7.23 | 2.93 | 0.9264 |

Table 2: Performance comparison of different models

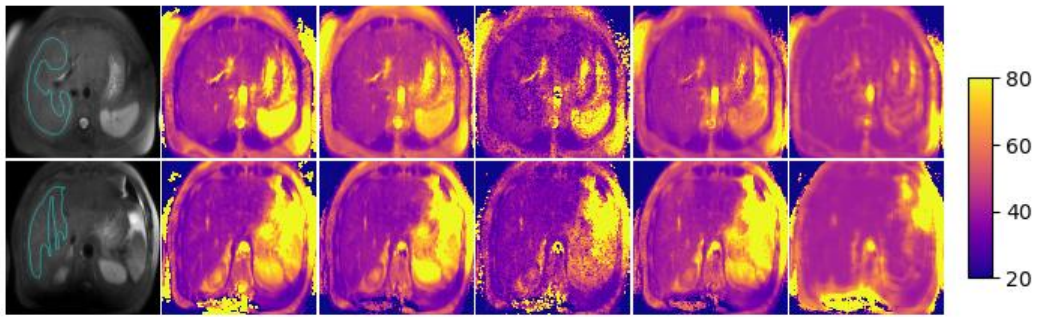

Figure 4: Examples of typical predicted $T_{1\rho}$ maps of two slices using various methods. From Left: Anatomy with ROI, Reference map, Proposed, 2-TSL, Two-stream, Supervised Learning. The unit is in millisecond.

**Fig. 5** shows the Bland-Altman Plot of the ROI mean value of different models in the comparison study. The ROI mean value of our proposed method demonstrates a decent agreement with the reference ROI mean value. The results of the Two-stream method also achieve a good agreement, while it has two outliers that are far away from the confidence interval. The *p* value of the hypothesis testing is 0.20>0.05(42.00±2.58 ms vs 42.20±3.94 ms), which did not reject the null hypothesis that there is no significant difference between the ROI mean value of the predicted maps from the proposed model and the reference maps.

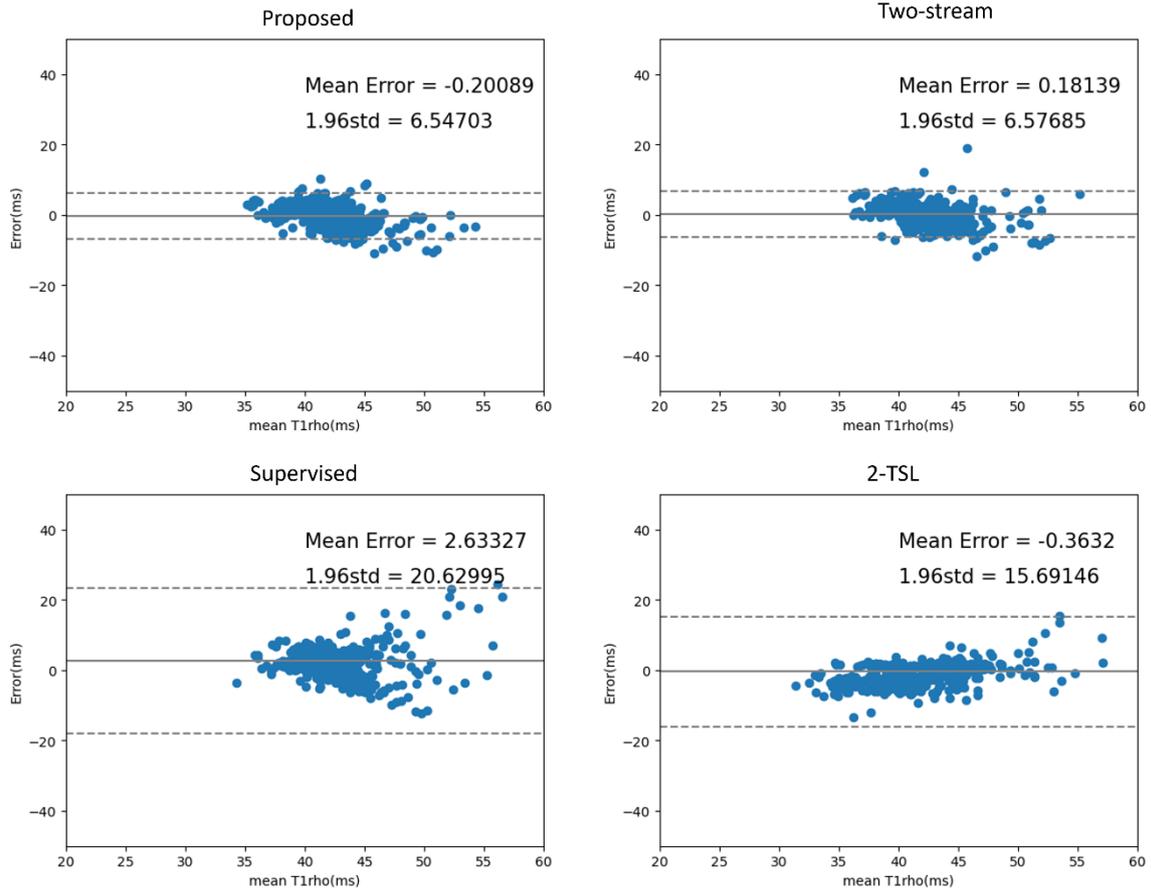

Figure 5. Bland-Altman plots of the ROI mean value of different models in the comparison study

## 3.3 Accuracy comparison of different combination of input

In our input setting, the input consisted of an image with $TSL = 0$ and an image with a nonzero $TSL$. Therefore, we compared how different nonzero $TSL$ images affected the model performance in predicting accuracy. The results are shown in **Table. 3**.

| RMAE(ms) | | | |
|---|---|---|---|
| Model | TSL = 10ms | TSL=30ms | TSL=50ms |
| 2-TSL | 11.39 | 8.69 | 2.28 |
| Two-stream | 5.68 | 3.62 | 3.53 |
| Supervised | 9.23 | 4.75 | 3.50 |
| Proposed | 3.84 | 3.54 | 3.44 |
| RMVAE(ms) | | | |
| Model | TSL = 10ms | TSL=30ms | TSL=50ms |
| 2-TSL | 6.71 | 3.10 | 1.41 |
| Two-stream | 5.06 | 2.91 | 2.87 |
| Supervised | 8.54 | 3.97 | 2.87 |
| Proposed | 2.97 | 2.94 | 2.80 |

Table 3: RMAE and RMVAE comparison with different TSL input. The unit is in millisecond

It can be seen that with the increase of $TSL$, the performance on accuracy increases. Our proposed method is more robust to the changes of the $TSL$ than the other methods. For example, though the 2-TSL method achieves the best performance at the longest $TSL$ (50 ms), it performs poorly at a shorter $TSL$. Discussion of these phenomenon is provided in the discussion section.

### 3.4 Ablation study

We also compared the following models for ablation study:

**Baseline model**: This model used the loss function in Eq [4], without dropout for epistemic uncertainty and without additional learning of the aleatoric uncertainty.

**Baseline model + epistemic**: This model used the loss function in Eq [4] with dropout enabled during inference for epistemic uncertainty but without additional learning of the aleatoric uncertainty.

**Baseline model + aleatoric**: This model used the loss function in Eq [8], without dropout for epistemic uncertainty but with additional learning for the aleatoric uncertainty.

**Baseline model + aleatoric + epistemic**: This is the proposed model with Bayesian uncertainty estimation.

**Table. 4** demonstrates the performance results of the ablated models. It can be seen that using the loss function in Eq [8](Baseline + aleatoric) can already achieve a considerable accuracy performance improvement by penalising the model from learning those noisy data, and the proposed method achieved the best results. It is also notable that applying dropout inference to the model can result in a slightly lower RSSIM, as the average computation of different forward pass may induce smoothing effect.

| Model | RMAE(ms) | RPSD(ms) | RMVAE(ms) | RSSIM |
|:---:|:---:|:---:|:---:|:---:|
| Baseline | 3.72 | 9.63 | 2.97 | 92.47 |
| Baseline +epistemic | 3.68 | 7.40 | 3.04 | 92.28 |
| Baseline + aleatoric | 3.62 | 7.56 | 2.99 | 93.02 |
| Proposed | 3.60 | 7.23 | 2.93 | 92.64 |

Table 4. Performance comparison of different models in the ablation study

**Fig. 6** shows the predicated $T_{1\rho}$ maps of different models in the ablation study. It can be seen that the Baseline model tends to overestimate the $T_{1\rho}$ values in the liver. By applying dropout and the uncertainty-aware loss function, the resulted maps become more comparable to the reference map.

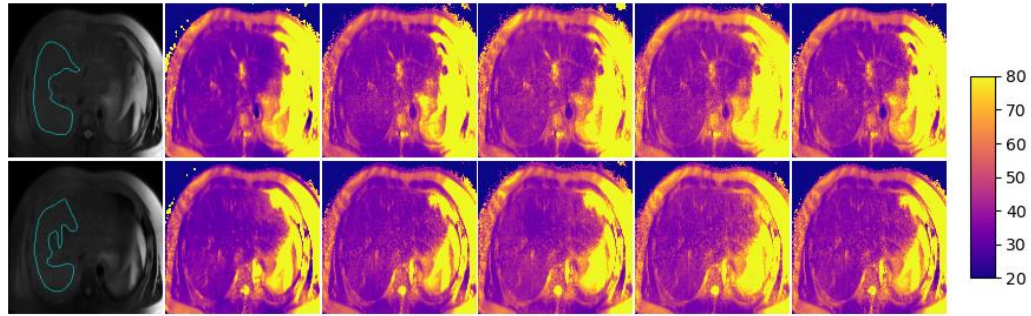

Figure 6. Examples of the predicted $T_{1\rho}$ maps of two slices using different models in the ablation study. From Left: Anatomy with ROI, Reference map, Proposed, Baseline, Baseline+Epistemic, Baseline+Aleatoric. The unit is in millisecond.

**3.5 Uncertainty Modelling Evaluation**

**Fig. 7** shows the sparsification plot of the pixels in the ROI of the four-fold cross-validation. The curve has a descending trend and it lies beneath the random curve, demonstrating a reasonable uncertainty modelling.

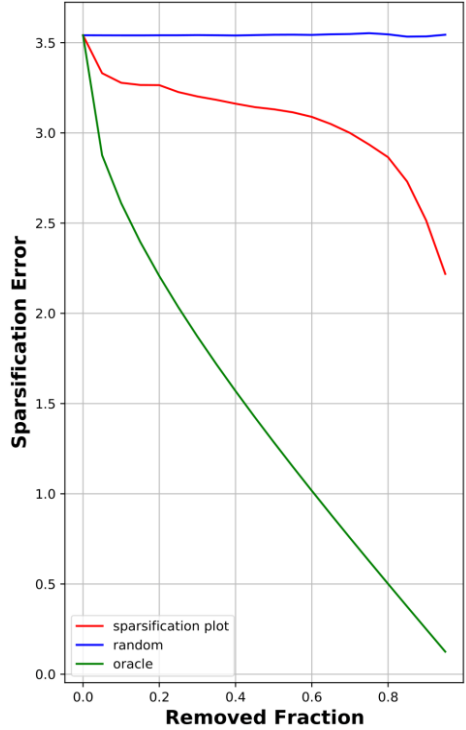

Figure 7. Sparsification plot on all the pixels in the ROIs from the four-fold validation results

**Table 5** shows the comparison results of the mean uncertainty value in the left and the right halves of the images of all four folds and the mean uncertainty value inside and outside the ROI. We observe that the mean uncertainty in the left half is lower than that in the right half, and the mean uncertainty inside the ROI is lower than that outside the ROI. This finding is consistent with the protocol setup in which signals near the liver and signals inside the ROI are more reliable, as those areas have less motion, and the $T_{1\rho}$ weighted signal is benefitted from applying the shim box and the suppression of the blood and fat signal.

| Left Half | Right Half | Inside ROI | Outside ROI |
|---|---|---|---|
| 3.25 | 3.74 | 3.21 | 3.65 |

Table 5: Uncertainty values of different areas

**Fig. 8** illustrates the examples of the uncertainty maps. The aleatoric uncertainty map and epistemic uncertainty map provided different information. The aleatoric uncertainty map highlighted areas where the acquired signal is likely to violate the mono-exponential physics model, and where it is difficult to make a precise $T_{1\rho}$ evaluation, as stated in Section 2.2. We observe that the stomachs on the right half of the image is highlighted because motion, flow effects and magnetic field inhomogeneity exist. Some blood vessels in the liver and the aorta are highlighted as they are with blood, motion and flow effects, and the $TSL$s used for liver scan do not guarantee reliable $T_{1\rho}$ quantification in these areas. The spines are highlighted because of cerebrospinal fluid. As for the epistemic uncertainty map, it mainly highlights the areas with image artefacts. For example, the banding in the vertical direction at the right edge of the image shown in the top row has high uncertainty. Small contexture structures with high frequency information are also highlighted. The convolutional neural network can have difficulties learning these features, as low-frequency features dominate the images (Ning et al., 2021, Ayyoubzadeh and Wu, 2021). Note the epistemic uncertainty is not obvious in many areas in the images since the testing dataset typically lacks abnormal liver information that is not seen in the training data.

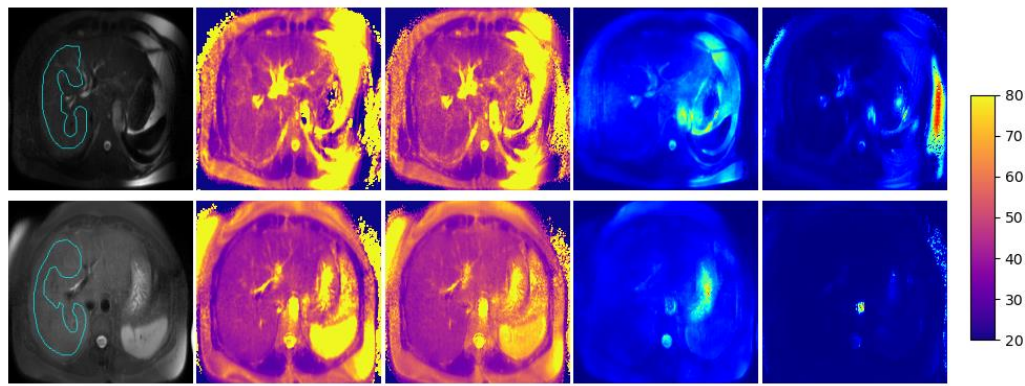

Figure 8: Examples of $T1\rho$ maps and the uncertainty maps. From Left: The anatomical image with ROI, the reference $T1\rho$ map, $T1\rho$ prediction, the aleatoric uncertainty map, the epistemic uncertainty map. All the uncertainty maps are normalized between 0 and 1. Best view in color and zoom in for details. The unit of the $T1\rho$ is in millisecond.

### 3.6 Uncertainty under abnormal perturbation

We added Gaussian noise to the images with different SNR levels and see if the mean uncertainty changes according to the changes of SNR. The results are shown in **Fig. 9.** We observe that the total uncertainty value decreases with an increase in SNR.

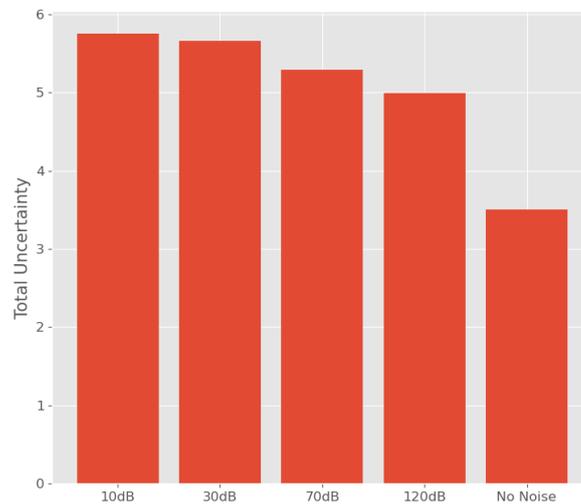

Figure 9: Changes of mean uncertainty value according to SNR by adding Gaussian noise.

We also conducted experiments to measure the uncertainty when the trained network is applied to predict $T_{1\rho}$ values on subjects with abnormal anatomy structures which was not seen during the training. Patterns or features that are not well-represented by the training data are expected to highlight in the epistemic uncertainty map. Specifically, we added a $20 \times 20$ birght square in the liver to test if the epistemic uncertainty map can highlight the area. We also examined the result on a tumour patient. Note this lesion has never been seen by the model in the training set. The results are shown in **Fig. 10** and **Fig .11**, respectively. The area with the abnormal square is highlighted, and the tumour has a relatively higher uncertainty value than the parenchyma. It is also noticeable **Fig.11** is taken at a location near the lower border of the liver, a location rarely used to acquire the training data set. Consequently, those non-liver structure which are rarely seen in the training data set such as the two kidneys are highlighted in the uncertainty map.

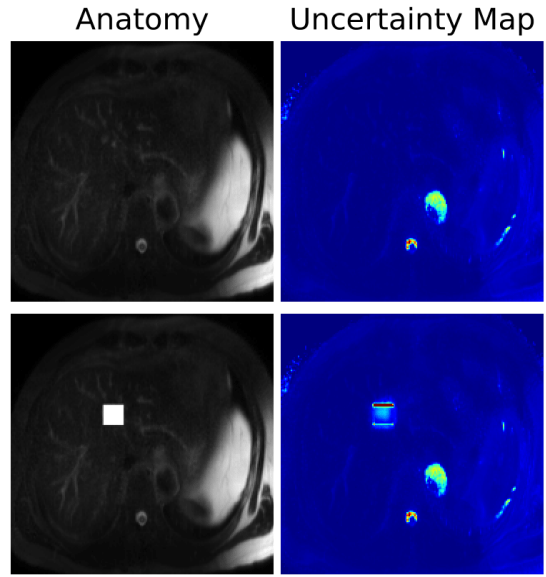

Figure 10: Epistemic uncertainty map before and after adding

a bright square

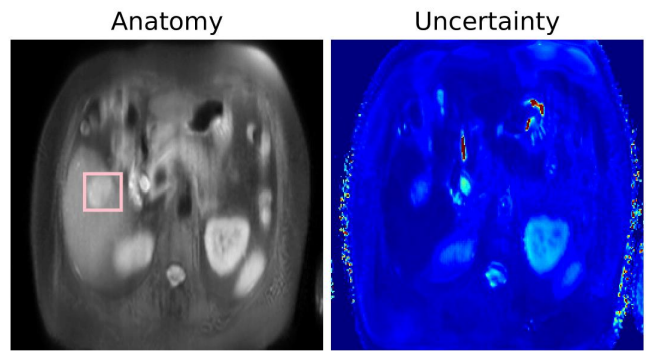

Figure 11: Epistemic uncertainty map of liver $T_{1\rho}$ mapping of a patient with liver tumor. The tumor is indicated with the pink box.

### 3.7 Explainability of the uncertainty awareness

We examine the explainability of the uncertainty awareness of the model by using the vanilla gradient saliency map (Simonyan et al., 2013) , which is a classic explainable AI tool. Specifically, we passed a two-channel test input into the network and computed the gradient of the loss function with respect to the input tensor. The saliency map is calculated to reflect the importance of every pixel by finding the pixel-wise maximum magnitude value of the gradient between the two channels. The results are shown in **Fig 12**. The model without aleatoric uncertainty produce saliency maps that have saliency points located outside the liver (i.e. the right half of the image) which has increased field inhomogeneities due to the application of the shimming box on the liver. In contrast, the saliency maps of the proposed method have considerably less saliency points at those areas. It is also shown that most of the saliency points spread at the left part of the image or at those areas with low aleatoric uncertainty. The result is aligned with the assumption that aleatoric uncertainty regularize the model to avoid using spatial information from areas with unreliable signal during training.

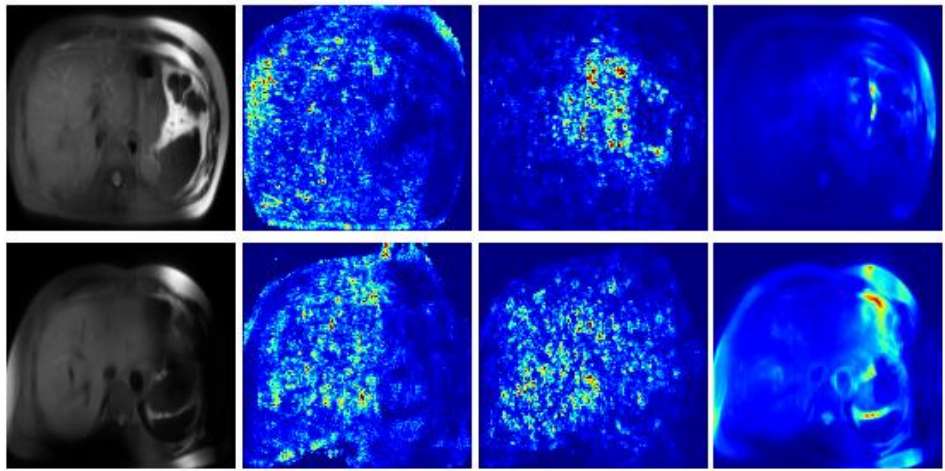

Figure 12: Vanilla gradient saliency results. From left to right: The anatomical images, the vanilla gradient saliency maps of the proposed model, the vanilla gradient saliency maps of the model "Baseline + Epistemic", the aleatoric uncertainty maps. The magnitude of the saliency maps is multiplied by three for a better visualisation.

## 4 Discussion and Limitations

The above results demonstrate the feasibility of learning a liver $T_{1\rho}$ mapping network without explicitly learning the constant $S_0$ map. By using the proposed relaxation constraint loss, the physics prior knowledge within the data is utilized in a compact way and this enforce the model to learn the $T_{1\rho}$ that satisfies the physics model.

It is also worth discussing why our model can map two $T_{1\rho}$ weighted images to a reasonable $T_{1\rho}$ prediction with not too much training data while the supervised learning produces inferior results. In fact, the relaxation constraint between $T_{1\rho}$ images in the dataset provides multiple supervision signal to regularise the network with physics prior knowledge. As we also adopt a training strategy that the loss of every constraint pair is back-propagated separately, there are multiple supervision signal for a single input pair. In other words, the relaxation relationship between images can be regarded as a kind of data augmentation. Conversely, directly applying the purely supervised learning method has limited supervision signal and regularisation. The only supervision comes from the $T_{1\rho}$ map fitted by four $T_{1\rho}$ images, and limited physics prior knowledge is used in the training. This could hamper the model's generality when data is limited and thus produce inferior results. The supervision signal from areas outside the liver parenchyma is also noisy and can further affect the performance.

It is interesting to discuss the phenomenon that the performance of the deep models improved as the $TSL$ prolonged. This can be explained from the same trend shown in the 2-TSL method. For a mono-exponential model, if the $TSL$ of the two $T_{1\rho}$ weighted images are close to each other, the final fitted $T_{1\rho}$ is highly susceptible to the disturbance of the signal. When it comes to the deep learning, the changes of image intensity between the two input images reflects the information encoded in the exponential model. The network's performance can be affected if the changes of image intensity deviate from the relaxation model due to image artifacts or noise. However, the deep model's results are still acceptable and more robust than the 2-TSL method since contextual information also contributes to the fitting of the parametric map in deep learning method. One might suggest that we can set a very long $TSL$ so that we can directly use the 2-TSL method to get promising results. However, long TSL leads to high RF energy deposition and tissue heating which is characterized by specific absorption rate (SAR). The maximum TSL is limited by the SAR allowed during scanning and the RF hardware. It is desirable to reduce SAR of the $T_{1\rho}$ imaging by reducing TSL. Therefore, the robust performance on images with short $TSL$ of our proposed deep model is also useful in practice.

It should also be pointed out that in this work we focused on on-resonance spin-lock with a simple mono-exponential relaxation model for $T_{1\rho}$ calculation. Other relaxation models, such as a mono-exponential model with a direct current component for on-resonance and off-resonance spin-lock (Jiang and Chen, 2018) or bi-exponential relaxation models, are worthy of further investigation.

As for the uncertainty estimation, we have demonstrated that the uncertainty map generally reflects the confidence level in liver $T_{1\rho}$ mapping, and it could potentially guide users to select the ROIs with reliable $T_{1\rho}$ quantification. It should also be pointed out that we use an indirect way to model the aleatoric uncertainty by minimising the distance between image signals. Previous works of other tasks such as self-supervised image registration and depth estimation have demonstrated that modelling aleatoric uncertainty in such an indirect way in image domain is feasible in reflecting the confidence trend and the task-specific domain knowledge qualitatively(Klodt and Vedaldi, 2018, Gong et al., 2022), which justifies ours. However, limitations of such an indirect way need to be addressed in the future. From a rigorous mathematical point of view, the aleatoric uncertainty is not strictly from the distribution of $T_{1\rho}$, as the image signal instead of $T_{1\rho}$ is directly minimised in the loss function. How to strictly estimate aleatoric uncertainty to characterize the inherent randomness of the imaging system during self-supervised learning remains to be explored. Another limitation is that we do not separate the contributions of different sources of data imperfections in modelling the uncertainty since these data imperfections are coupled together in acquired images. It is worthy of further study to understand the contribution from each source on data uncertainty. In this work, we manually draw ROIs to perform the analysis. Methods which can be used for automatic ROI selection such as those based on attention mechanism (Niu et al., 2021) can be useful in future studies. The uncertainty map itself may be also used to aid the automatic selection of ROIs, as the uncertainty map indicate the trustworthiness of different areas in the quantitative map. Lastly, our data sets are collected at a single site with the same MRI protocol. Future multi-sites independent studies and research on the application of domain adaptation techniques(Guan and Liu, 2021) in qMRI is beneficial for general use of learning-based qMRI system.

## 5 Conclusion

Our proposed relaxation-constrained self-supervised learning loss can enable a deep model to provide $T_{1\rho}$ quantification using two $T_{1\rho}$ weighted images. Our Bayesian $T_{1\rho}$ quantification network approximates both aleatoric and epistemic uncertainty in addition to $T_{1\rho}$ quantification. Our experiments demonstrated the proposed method outperforms the existing methods in $T_{1\rho}$ quantification of the liver, and the estimated uncertainty accord with the confidence level of $T_{1\rho}$ in liver imaging. Future work includes modelling the $T_{1\rho}$ uncertainty from a more rigorous perspective and relaxation-constraints using other relaxation models, as well as conducting multi-site study.

**Acknowledgement**

This study was supported by a grant from the Research Grants Council of the Hong Kong SAR (Project GRF 14201721), a grant from the Innovation and Technology Commission of the Hong Kong SAR (Project No. MRP/046/20X), a grant from the Hong Kong Health and Medical Research Fund (HMRF) 06170166, and a grant from the Research Grants Council of the Hong Kong SAR (Project SEG No. CUHK02).